\documentclass[aps,prl,letterpaper,amsmath,amssymb,superscriptaddress,reprint]{revtex4-1}
\usepackage{graphicx}

\begin{document}
\title{Unconventional superconductivity in YPtBi and related topological semimetals}
\author{Markus Meinert}
\email{meinert@physik.uni-bielefeld.de}
\affiliation{Center for Spinelectronic Materials and Devices, Bielefeld University, D-33501 Bielefeld, Germany}
\date{\today}

\begin{abstract}
YPtBi, a topological semimetal with very low carrier density, was recently found to be superconducting below $T_\mathrm{c} = 0.77$\,K. In the conventional theory, the nearly vanishing density of states around the Fermi level would imply a vanishing electron-phonon coupling and would therefore not allow for superconductivity. Based on relativistic density functional theory calculations of the electron-phonon coupling in YPtBi it is found that carrier concentrations of more than $10^{21}$\,cm$^{-3}$ are required to explain the observed critical temperature with the conventional pairing mechanism, which is several orders of magnitude larger than experimentally observed. It is very likely that an unconventional pairing mechanism is responsible for the superconductivity in YPtBi and related topological semimetals with the Half-Heusler structure.
\end{abstract}

\maketitle

A series of Half-Heusler compounds with heavy elements were predicted to have topologically non-trivial band order \cite{Lin2010,Al-Sawei2010, Feng2010}. These compound have high cubic symmetry, however without inversion symmetry. The normal band order with the $s$-like, twofold degenerate $\Gamma_6$ state sitting above the $p$-like, fourfold degenerate $\Gamma_8$-state is inverted in some of these compounds due to spin-orbit coupling. In their natural state, the cubic symmetry leads to a band degeneracy at $\Gamma$ around the Fermi level, rendering them semimetals with topologically non-trivial band order (topological semimetals) and very low density of states (DOS) at the Fermi level  $D(E_\mathrm{F})$. By breaking the cubic symmetry with some amount of uniaxial strain, the compounds can be made insulating, so they could become 3D topological insulators \cite{Qi2011}. They would exhibit metallic surface states with Dirac-like dispersion, i.e., the electrons behave as massless particles with ultrahigh mobility, while at the same time being insulating in the bulk. These surface states are topologically protected as long as time-reversal symmetry is preserved, i.e., they are protected against scattering from non-magnetic impurities. Indeed, for some of these materials there is experimental evidence for topologically nontrivial bandstructures and the presence of Dirac surface states \cite{Liu2011, Shekhar2012, Nowak2015}, although none were found to be insulating in the bulk. Some compounds from this class were found to be superconductors with critical temperatures up to 1.8\,K, e.g. LaPtBi, LuPtBi, LuPdBi, YPtBi, YPdBi \cite{Goll2008,Shekhar2013,Tafti2013,Xu2014,Nakajima2015}. Compounds of the type $R$PdBi ($R$ is a lanthanide with an open $4f$ shell) that show coexisiting local moment antiferromagnetism as well as superconductivity were found, pointing to the presence of spin triplet Cooper pairs \cite{Nakajima2015}, which is allowed due to the missing structural inversion symmetry \cite{Sigrist2007}. Due to the topologically nontrivial band structure, novel collective excitations are possible, in particular surface Majorana fermions \cite{Sato2009}. These could provide the basis for low-decoherence quantum processing \cite{Leijnse2012}.

Superconducting semiconductors such as GeTe and SnTe are long known \cite{Hein1964,Allen1969} and their superconductivity can be explained \cite{Allen1969} with the Eliashberg theory of electron-phonon mediated superconductivity. SrTiO$_{3-x}$, the most dilute semiconductor known to date, has $T_\mathrm{c} \lesssim 0.5\,\mathrm{K}$ and its superconductivity can not be explained by simple electron-phonon coupling \cite{Schooley1965, Bustarret2015}. Instead, a plasmon-assisted mechanism was proposed to explain the unusual dependence of $T_\mathrm{c}$ on the carrier density \cite{Takada1980}. From the BCS theory the well-known expression for the superconducting transition temperature $k_\mathrm{B} T_\mathrm{c} = 1.13 \, \hbar \omega_\mathrm{c} \, \exp{\left[ - 1/D(E_\mathrm{F}) V \right]}$ is obtained, where $\omega_\mathrm{c}$ is a cutoff frequency, that is often identified with the Debye frequency, and $V$ is the effective interaction potential. Thus, the critical temperature is expected to increase with increasing $D(E_\mathrm{F})$. This expectation was confirmed for GeTe and SnTe, but in these materials $T_\mathrm{c}$ is limited to a few hundred mK \cite{Allen1969}. More recently, superconductivity just below $T_\mathrm{c} \approx 4\,\mathrm{K}$ was discovered in highly B doped diamond \cite{Ekimov2004} and in Cu$_x$Bi$_2$Se$_3$ \cite{Hor2010}, a prototype topological insulator for $x=0$. For the latter case, it was shown by first principles calculations of the electron-phonon coupling that the conventional pairing mechanism is most likely not strong enough to give rise to the rather high observed critical temperature \cite{Zhang2015}. Furthermore, the possibility of superconductivity in the surface states of topological insulators was explored recently \cite{DasSharma2013, Li2014}, and it was shown that surface electron-phonon interaction can be strong.

Naturally the question arises whether the superconductivity in the topological Half-Heusler compounds is of the conventional, phonon-mediated type. To shed some light on this question, first principles calculations of the electron-phonon coupling for a topological semimetal from the Half-Heusler class are presented here. The compound YPtBi was chosen as a representative member of this class, which is a bulk superconductor with a critical temperature of $T_\mathrm{c} = 0.77$\,K \cite{Butch2011, Bay2012, Shekhar2013, Bay2014}.

The calculations were carried out with the \textsc{Quantum Espresso} distribution \cite{QE} within the framework of density functional theory (DFT). Relativistic PAW potentials \cite{DalCorso2010} (including spin-orbit coupling, SOC) from the PSlibrary \cite{DalCorso2014} were employed with kinetic energy cutoffs of 40\,Ry for the wavefunctions and 400\,Ry for the charge density. The results were checked against all-electron calculations including SOC with the full potential linearized augmented plane-wave (FLAPW) method with the \textsc{elk} code \cite{elk}  and were found to agree very well. The Perdew-Burke-Ernzerhof (PBE) generalized gradient approximation (GGA) was used for the exchange-correlation energy and potential. Since YPtBi is a semimetal, the use of a semilocal potential to describe the band structure is justified. The dynamical matrices and electron-phonon matrix elements $g_{\mathbf{k} + \mathbf{q}, \mathbf{k}}^{\mathbf{q}\nu, mn}$ with phonon mode index $\nu$ and band indices $m,n$ were obtained with density functional perturbation theory on a $5 \times 5 \times 5$ $\mathbf{q}$-point mesh and $10 \times 10 \times 10$ $\mathbf{k}$-point mesh. Electron-phonon coupling was calculated applying an interpolation scheme described in Ref. \onlinecite{Wierzbowska2005}. The Eliashberg spectral function
\begin{align}
\alpha^2 F(\omega) = &\frac{1}{D(\varepsilon_\mathrm{F})} \sum\limits_{mn}\sum\limits_{\mathbf{q}\nu}\sum\limits_\mathbf{k} \delta(\omega - \omega_{\mathbf{q}\nu})\nonumber
 \left|g_{\mathbf{k} + \mathbf{q}, \mathbf{k}}^{\mathbf{q}\nu, mn} \right|^2\nonumber\\
&\times \delta(\varepsilon_{\mathbf{k}+\mathbf{q},m} - E_\mathrm{F}) \delta(\varepsilon_{\mathbf{k},n} - E_\mathrm{F})
\end{align}
was evaluated on $40^3$ $\mathbf{k}$-point and $20^3$ $\mathbf{q}$-point meshes. The spectral function was obtained for several widths of a Gaussian approximation for the $\delta$-function between 0.004 and 0.02\,Ry. The resulting density of states $D(E_\mathrm{F})$ and electron-phonon coupling constant $\lambda$, where
\begin{equation}
\lambda = 2 \int \frac{\alpha^2 F(\omega)}{\omega} \mathrm{d}\omega,
\end{equation}
 were extrapolated with a linear fit to the $\delta \to 0$ limit. Convergence tests suggest that the accuracy for $\lambda$ is of the order $\pm 0.02$. The superconducting gap $\Delta$ was calculated by solving the isotropic Eliashberg equations self-consistently as a function of temperature with a routine implemented in the \textsc{elk} code \cite{elk}, which is based on a similar algorithm as described in Ref. \onlinecite{Margine2013}. The critical temperature $T_\mathrm{c}$ was obtained by interpolating $\Delta(T)$ with cubic splines and finding the inflection point. Additionally, the McMillan-Allen-Dynes formula
\begin{equation}\label{eq:McMillan-Allen-Dynes}
T_\mathrm{c} = \frac{\omega_\mathrm{log}}{1.2} \exp{\left[ - \frac{1.04 (1+\lambda)}{\lambda - \mu^* (1+0.62\lambda)} \right]}.
\end{equation}
with the logarithmic average phonon frequency $\omega_\mathrm{log}$ and the screened Coulomb pseudopotential $\mu^*$ was evaluated for comparison \cite{Allen1975}. The Coulomb pseudopotential is usually taken as a parameter of the order $\mu^* = 0.1 \dots 0.2$, which we set to zero for the evaluation of the Eliashberg equations and the McMillan-Allen-Dynes equation, i.e. Coulomb repulsion is completely neglected. Thus, the critical temperatures obtained here are upper bounds for the electron-phonon induced superconductivity. Doping was treated in a rigid-band approximation by adding or subtracting electrons from the band structure and compensating for this by adding a homogeneous background charge. The experimental lattice constant of 6.65\,\AA{} was used in all calculations.  In systems with very low Fermi energy ($E_\mathrm{F} \lesssim \omega_\mathrm{log}$), vertex corrections should in principle be included, which were shown to increase $T_\mathrm{c}$ to some extent \cite{Takada1993}. However, for all carrier densities considered here we have $E_\mathrm{F} \gg \omega_\mathrm{log}$, so the Eliashberg equations are expected to properly describe the electron-phonon interaction in YPtBi.

\begin{figure}[t]
\includegraphics[width=8.6cm]{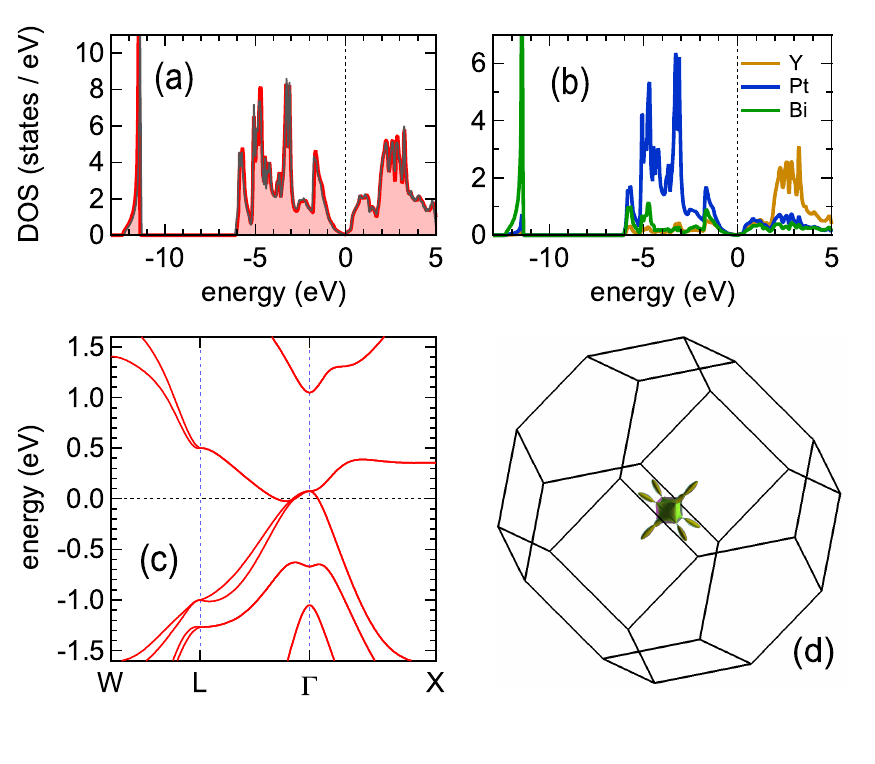}
\caption{\label{fig:dos_bands} (a) Density of states (DOS), (b) atom-projected density of states (PDOS), (c) band structure, and (d) Fermi surface of YPtBi obtained from FLAPW calculations. DOS plots obtained from the relativistic PAW (thin gray) and FLAPW (thick red) calculations are compared in (a) to show the equivalence of the two methods.}
\end{figure}

For an accurate description of the electron-phonon coupling, we need to make sure that the calculated band structure close to $E_\mathrm{F}$ is in agreement with the observed one. In the following, the computed electronic properties will be compared to experimental data to assess the validity of the band structure calculations. In Figure \ref{fig:dos_bands} the DOS, atom-projected DOS, band structure along high-symmetry lines and the Fermi surface are plotted. The relativistic PAW and FLAPW calculations are in excellent agreement, establishing that the PAW method with the potentials from the PSlibrary gives reliable results for YPtBi. The band structure plot clarifies that YPtBi has weakly overlapping bands close to the Brillouin zone center, making it a semimetal. The Fermi surface consists of two hole pockets and two sets of electron pockets, which belong to the fourfold degenerate $\Gamma_8$ representation \cite{Feng2010}. As shown in Ref. \cite{Feng2010}, the $\Gamma_6$ states lie below $\Gamma_8$, so that the band order is topologically nontrivial. The hole pockets are of approximately cubic shape centered around the zone center with corners along the $\Lambda$ path. The inner hole pocket has a complex concave-convex shape and touches the outer hole pocket on the $\Lambda$ and $\Delta$ paths. The electron pockets form a set of cigar-shaped ellipsoids with eightfold symmetry along the $\Lambda$ paths. The small Fermi surface is in agreement with the small density of states at the Fermi energy of $D(E_\mathrm{F}) = 0.038\,\mathrm{states}/\mathrm{eV}$. The calculated $D(E_\mathrm{F})$ agrees very well with the value obtained from a heat capacity measurement,  $D(E_\mathrm{F}) \lesssim 0.042\,\mathrm{states}/\mathrm{eV}$ \cite{Pagliuso1999}.

The Fermi surface obtained for YPtBi is remarkably similar to the Fermi surface of LaPtBi \cite{Oguchi2001}. The Fermi vector in $[001]$ direction is $k_\mathrm{F}^{[001]} \approx  0.033\,a_0^{-1}$ and the volume enclosed by the two hole pockets is approximately $V_\mathrm{F}^h \approx 0.000575\,a_0^{-3}$. The corresponding carrier density is $n_e = n_h = 3 \cdot 10^{19}\,\mathrm{cm}^{-3}$. Recently, Shubnikov-de Haas (SdH) oscillations in YPtBi single crystals with the magnetic field along the $[001]$ direction were observed with a frequency of $\Delta B^{-1} = 0.022\,\mathrm{T}^{-1}$ \cite{Butch2011}, indicating that the true Fermi surface cross section is even three times smaller than the calculated one. For the latter, a frequency of SdH oscillations with a periodicity of $\Delta B^{-1}_{[001]} \approx 0.0061\,\mathrm{T}^{-1}$ is expected. Thus, the dip of the conduction band minimum below the valence band maximum on the $\Lambda$ path is in fact not as deep as calculated. A beating node in the measurement indicates that there are two similar-sized Fermi surfaces contributing to the SdH oscillations, in perfect agreement with the band structure calculation. Simultaneous Hall effect measurements were analyzed with a one-band model giving $n_h = 2 \cdot 10^{18}\,\mathrm{cm}^{-3}$. Based on the observed cross section of the Fermi surface and assuming cubic shape, one obtains an enclosed charge density of $n_h \approx 4 \cdot 10^{18}\,\mathrm{cm}^{-3}$, which agrees nicely with the Hall effect measurement, however neglecting multiband effects, different mobilities and carrier compensation. This result supports that the true Fermi surface is much smaller than the calculated one. The surface-averaged effective masses on the calculated Fermi surface are $m_\mathrm{h}^* = 0.2\,m_\mathrm{e}$ and $m_{e}^* = 0.41\,m_\mathrm{e}$ for the hole and electron pockets, respectively, where $m_\mathrm{e}$ is the free-electron mass. The hole effective mass is roughly in agreement with the value extracted from SdH oscillation, $m_h^* = 0.15\,m_\mathrm{e}$ \cite{Butch2011}. In conclusion, the DFT electronic structure calculation with the PBE functional reproduces the experimental observations very well up to a small error in the overlap between conduction and valence bands. Thus,  the calculated electronic structure provides a solid foundation for the evaluation of the electron-phonon coupling discussed in the following.

\begin{figure}[t]
\includegraphics[width=8.6cm]{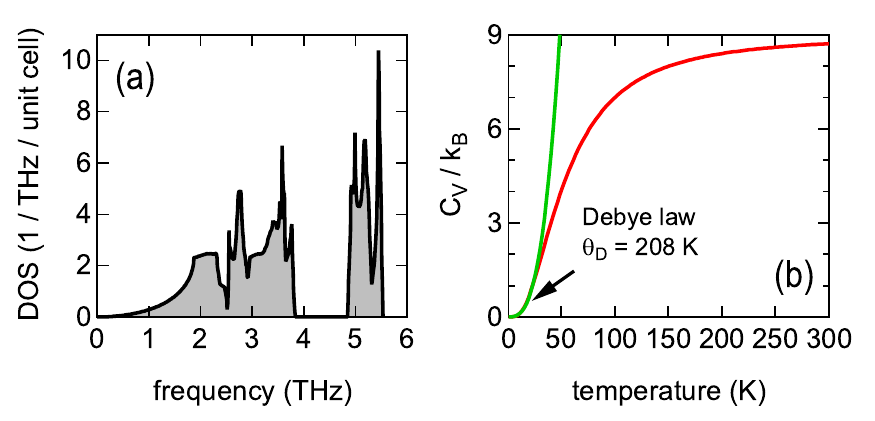}
\caption{\label{fig:phonon_dispersion}  Phonon density of states (a) and derived heat capacity (b) of pristine YPtBi.}
\end{figure}

The phonon density of states and heat capacity of YPtBi are shown in Figure \ref{fig:phonon_dispersion}. A clear separation of acoustic and optical modes is visible. From the low-temperature part of the heat capacity a Debye temperature of $\Theta_\mathrm{D} = 208\,\mathrm{K}$ is obtained, in good agreement with the experimental value of $195(5)\,\mathrm{K}$ \cite{Pagliuso1999}. Because of the small Fermi surface, only very short $\mathbf{q}$-vectors with $q \leq 2k_\mathrm{F}$ can connect different parts of the surface, which gives rise to the scattering of an electron state into another. Thus, the electron-phonon coupling is limited to a small region close to the zone center. From the extrapolation scheme for the Brillouin zone integration $\lambda = 0.02 \pm 0.02$ is obtained, obviously at the limit of the numerical accuracy of the Brillouin zone sampling. Certainly $\lambda$ is small, but a more accurate $\lambda$ will require an unfeasibly dense $\mathbf{q}$-point mesh. With $\mu^* = 0$ the solution of the Eliashberg equations shows that $T_\mathrm{c} < 0.001$\,K, much smaller than the observed critical temperature of $T_\mathrm{c} = 0.77\,\mathrm{K}$.

\begin{figure}[t]
\includegraphics[width=8.6cm]{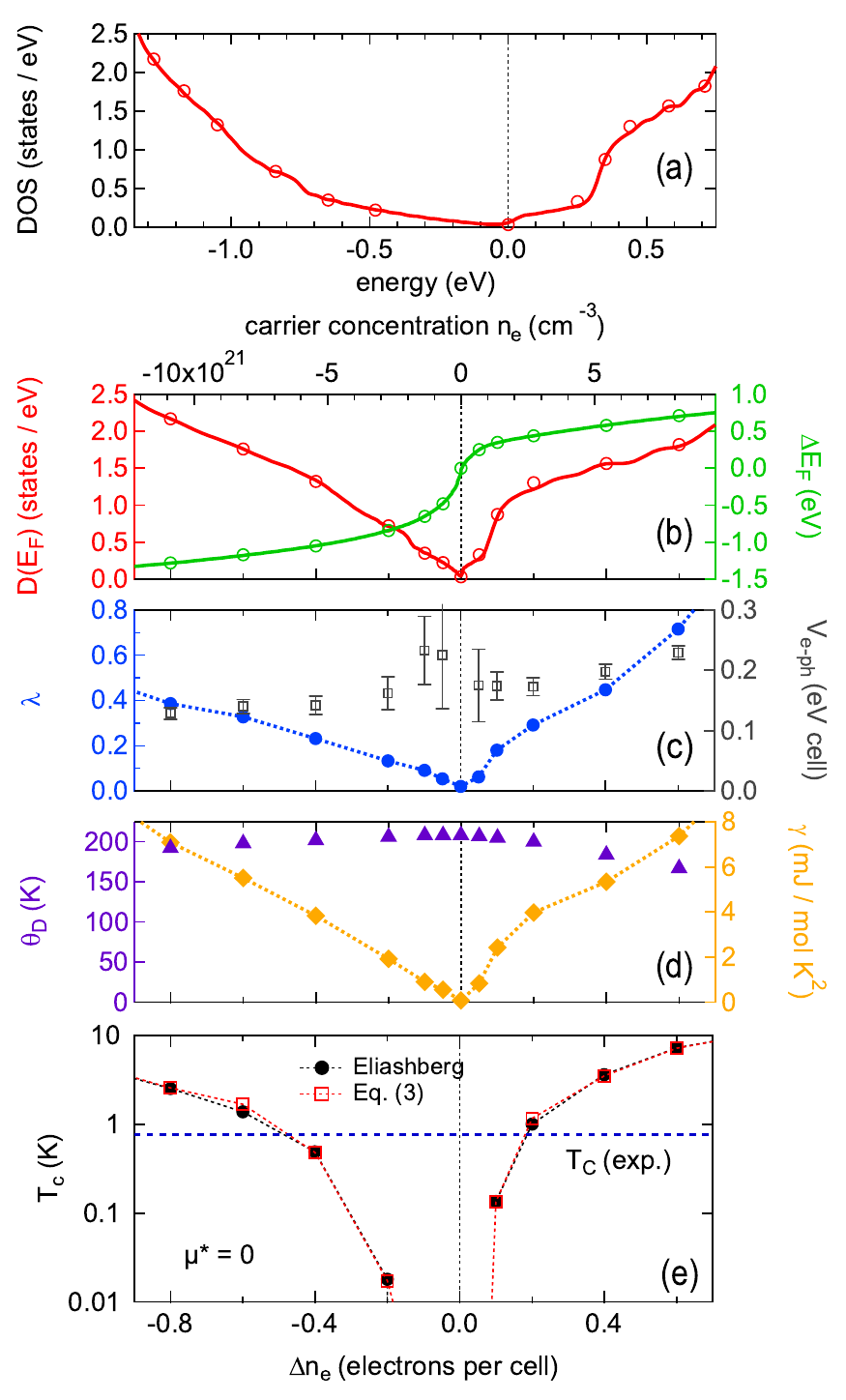}
\caption{\label{fig:electron_phonon_results}(a) Density of states as a function of the energy. Open circles mark the values for which electron-phonon calculations were done. (b) Density of states and Fermi energy shift as a function of additional electrons per primitive cell $\Delta n_\mathrm{e}$. (c) and (d) Electron-phonon coupling constant $\lambda$, interaction potential  $V_\mathrm{e-ph} = \lambda / (1 + \lambda) \cdot D(E_\mathrm{F})^{-1}$, Debye temperature $\Theta_\mathrm{D}$ and Sommerfeld coefficient $\gamma$ as functions of $\Delta n_\mathrm{e}$. (e) Calculated critical temperatures from numerical solution of the Eliashberg equations and from the McMillan-Allen-Dynes formula.}
\end{figure}

To investigate the effect of doping, electron-phonon coupling was evaluated for doping levels of $\Delta n_e = \pm 1.0$ electrons per primitive cell. The dynamical matrices were recomputed for each doping level, so that phonons  were treated at full self-consistency with respect to doping. The corresponding densities of states and Fermi energy shifts are given in Fig. \ref{fig:electron_phonon_results} (a) and (b). Because of the low DOS close to $E_\mathrm{F}$ in the undoped YPtBi, small doping levels already give rise to large Fermi energy shifts. With increasing doping of both electron- or hole-type, $T_\mathrm{c}$ is increased as seen in Fig. \ref{fig:electron_phonon_results} (e), however electron doping ($\Delta n_e > 0$) is clearly more effective. The coupling leads to a renormalization of the phonon frequencies, which is indicated by the reduction of the Debye temperature, Fig. \ref{fig:electron_phonon_results} (d). This also indicates that YPtBi could be dynamically unstable at strong doping. 

The observed critical temperature of $T_\mathrm{c} = 0.77\,\mathrm{K}$ is obtained at $\Delta n_e \approx +0.2 = 2.8 \times 10^{21}$\,cm$^{-3}$ or $\Delta n_e \approx -0.5 = 7 \times 10^{21}$\,cm$^{-3}$ with $\mu^* = 0$, see Fig. \ref{fig:electron_phonon_results} (e). These concentrations are lower bounds, since $T_\mathrm{c}$ decreases with $\mu^* > 0$. As seen in Fig. \ref{fig:electron_phonon_results} (e), the McMillan-Allen-Dynes formula closely resembles the full numerical solution of the Eliashberg equations and remains valid down to $T_\mathrm{c} \approx 0.02$\,K. In all cases, the BCS expression for the superconducting gap, $2 \Delta(0) = 3.528 k_\mathrm{B} T_\mathrm{c}$ is approximately fulfilled by the numerical solutions. Also, the usual relation of $\lambda$ with the renormalization function $Z(\omega=0) = 1 + \lambda$ was found to be fulfilled in all cases, demonstrating the consistency of the Eliashberg calculations.

To study the influence of $D(E_\mathrm{F})$, the electron-phonon interaction potential $V_\mathrm{e-ph} = \lambda / (1 + \lambda) \cdot D(E_\mathrm{F})^{-1}$ is given in Fig. \ref{fig:electron_phonon_results} (c). Over the full range of doping concentrations the interaction potential is $V_\mathrm{e-ph} \approx 0.2\,\mathrm{eV\,cell}$, so the low value of $\lambda$ for weakly and undoped YPtBi comes mainly from the low $D(E_\mathrm{F})$, or, equivalently, from the small Fermi surface area. These doping levels could be realized through off-stoichiometry (YPtBi crystals are mostly grown out of Bi flux, so additional Bi could easily be incorporated), or locally due to site-swap between neighboring cells. Also grain boundaries and other inhomogeneities with different stoichiometry could serve as sources of intrinsic doping. However, the carrier concentrations required for the electron-phonon coupling to be strong enough to explain the observed critical temperature are at least one order of magnitude larger than the typically observed carrier concentrations in YPtBi and three orders of magnitude larger than for the samples with lowest observed carrier concentration ($2 \times 10^{18}$~cm$^{-3}$, Ref. \onlinecite{Butch2011}). Because of the increase in $D(E_\mathrm{F})$ the Sommerfeld coefficient of the heat capacity, $\gamma = \pi^2D(E_\mathrm{F}) (1+\lambda) k_\mathrm{B}^2 / 3$ would increase to values around $\gamma \approx 4\,\mathrm{mJ/mol\,K^2}$ (see Fig. \ref{fig:electron_phonon_results} (d)), much larger than the measured value $\gamma \lesssim 0.1\,\mathrm{mJ/mol\,K^2}$ \cite{Pagliuso1999}. Experiments with high-quality samples indicate that the normal-state electronic properties of YPtBi are perfectly in agreement with the calculation for the ideal, undoped case \cite{Butch2011, Bay2012}. A diamagnetic screening fraction around 70\% was observed in the superconducting state, underlining that a large part of the material is in the superconducting state and thereby ruling out the possibility of grain-boundary superconductivity \cite{Bay2014} or surface superconductivity \cite{DasSharma2013}. The calculated critical temperature in the experimentally observed carrier concentration range of $2 \times 10^{18}$\,cm$^{-3}$ to $4.2 \times 10^{20}$\,cm$^{-3}$  \cite{Butch2011, Bay2012, Shekhar2013, Bay2014} is $T_\mathrm{c} \ll 0.1$\,K. However, an even more remarkable observation is that the experimental critical temperature of YPtBi varies little across different samples despite the carrier concentrations vary over two orders of magnitude. From Fig. \ref{fig:electron_phonon_results} (e) one would expect a variation of $T_\mathrm{c}$ over several orders of magnitude in that range.

The lack of a clear correlation between normal-state electronic properties, sample quality, and critical temperature indicates that electron-phonon interaction induced by doping is not an explanation for the superconductivity in YPtBi. The relation between critical field and temperature observed in Ref. \onlinecite{Bay2012} deviates from conventional $s$-wave behavior and suggests that the material could be a $p$-wave superconductor, very similar to Cu$_x$Bi$_2$Se$_3$. On the other hand, Cooper pair wave functions with angular orbital momentum $l > 0$ are not protected by the Anderson theorem, so random scattering from defects and impurities should reduce $T_\mathrm{c}$ if the elastic mean-free-path $\ell$ is smaller than the superconducting coherence length $\xi$ \cite{Mackenzie2003}. Values of $\xi = 15, 17$\,nm and $\ell = 105, 130$\,nm were observed for YPtBi \cite{Butch2011, Bay2012}, but superconductivity was also reported in a case where the mean free path based on free-electron theory, i.e.  $\ell = \hbar k_\mathrm{F} / \rho_0 n e^2$ with $k_\mathrm{F} = (3\pi^2 n)^{1/3}$ is smaller than the lattice constant \cite{Shekhar2013}. This indicates that YPtBi is superconducting even in the dirty limit, which apparently contradicts the hypothesis of $p$-wave superconductivity. A remarkable side-note is that many Half-Heusler compounds are known as semiconductors that show large thermoelectric power at appropriate doping. Even though the carrier concentrations are often high, superconductivity has never been reported for any of these compounds \cite{Graf2011}.

Based on the analysis of the electron-phonon coupling and comparison with experimental data on the normal-state properties it is safe to conclude that an unconventional mechanism is responsible for the superconductivity in YPtBi. Related compounds from the class of topological Half-Heusler semimetals, $R$PtBi and $R$PdBi (with a rare-earth element $R$) have very similar normal-state properties as YPtBi and also show superconductivity with critical temperatures up to 1.8\,K. It is most likely that an unconventional pairing mechanism is at work in all of these compounds. More experimental work, in particular careful studies on the interplay between structural order, normal state electronic properties and superconductivity are necessary to gain more systematic knowledge about the pairing mechanism and the parity of the Cooper pairs. From the theoretical point of view it is particularly challenging to identify pairing mechanisms that allow for sufficiently strong coupling despite the low Fermi density of states, for example an electron-electron coupling assisted by plasmons \cite{Takada1980}.

\acknowledgments
Calculations leading to the results presented here were performed in part on resources provided by the Paderborn Center for Parallel Computing. The author thanks Thomas Dahm for fruitful discussions.


\begin{thebibliography}{50}

\bibitem{Lin2010} H. Lin, L. A. Wray, Y. Xia, S. Xu, S. Jia, R. J. Cava, A. Bansil, and M. Z. Hasan, Nat. Mater. \textbf{9}, 546 (2010).
\bibitem{Al-Sawei2010} W. Al-Sawai, H. Lin, R. S. Markiewicz, L. A. Wray, Y. Xia, S. Y. Xu, M. Z. Hasan, and A. Bansil, Phys. Rev. B \textbf{82}, 125208 (2010).
\bibitem{Feng2010} W. Feng, D. Xiao, Y. Zhang, and Y. Yao, Phys. Rev. B \textbf{82}, 235121 (2010).
\bibitem{Qi2011} X.-L. Qi and S.-C. Zhang, Rev. Mod. Phys. 83, 1057 (2011).
\bibitem{Liu2011}C. Liu, Y. Lee, T. Kondo, E. D. Mun, M. Caudle, B. N. Harmon, S. L. Budko, P. C. Canfield, and A. Kaminski, Phys. Rev. B \textbf{83}, 205133 (2011).
\bibitem{Shekhar2012} C. Shekhar, S. Ouardi, A. K. Nayak, G. H. Fecher, W. Schnelle, and C. Felser, Phys. Rev. B \textbf{86}, 155314 (2012).
\bibitem{Nowak2015} B. Nowak, O. Pavlosiuk, and D. Kaczorowski, J. Phys. Chem. C 119, 2770 (2015).

\bibitem{Goll2008} G. Goll, M. Marz, A. Hamann, T. Tomanic, K. Grube, T. Yoshino, and T. Takabatake, Physica B: Condens. Matt. \textbf{403}, 1065 (2008). 
\bibitem{Shekhar2013} C. Shekhar, M. Nicklas, A. K. Nayak, S. Ouardi, W. Schnelle, G. H. Fecher, C. Felser, and K. Kobayashi, J. Appl. Phys. \textbf{113}, 17E142 (2013).

\bibitem{Tafti2013} F. F. Tafti, T. Fujii, A. Juneau-Fecteau, S. Rene de Cotret, N. Doiron-Leyraud, A. Asamitsu, and L. Taillefer, Phys. Rev. B 87, 184504 (2013).
\bibitem{Xu2014} G. Xu, W. Wang, X. Zhang, Y. Du, E. Liu, S. Wang, G. Wu, Z. Liu, and X. X. Zhang, Sci. Rep. 4, 5709 (2014). 
\bibitem{Nakajima2015} Y. Nakajima, R. Hu, K. Kirshenbaum, A. Hughes, P. Syers, X. Wang, K. Wang, R. Wang, S. R. Saha, D. Pratt, J. W. Lynn, and J. Paglione, Sci. Adv. 1500242 (2015). 
\bibitem{Sigrist2007} M. Sigrist, D. F. Agterberg, P. A. Frigeri, N. Hayashi, R. P. Kaur, A. Koga, I. Milat, K. Wakabayashi, and Y. Yanase, J. Magn. Magn. Mater. 310, 536 (2007).
\bibitem{Sato2009} M. Sato and S. Fujimoto, Phys. Rev. B 79, 094504 (2009).
\bibitem{Leijnse2012} M. Leijnse and K. Flensberg, Semicond. Sci. Technol. \textbf{27}, 124003 (2012).
\bibitem{Hein1964} R. A. Hein, J. W. Gibson, R. Mazelsky, R. C. Miller, and J. K. Hulm, Phys. Rev. Lett. 12, 320 (1964).
\bibitem{Allen1969} P. B. Allen and M. L. Cohen, Phys. Rev. 177, 704 (1969).
\bibitem{Schooley1965} J. F. Schooley, W. R. Hosler, E. Ambler, J. H. Becker, M. L. Cohen, and C. S. Koonce, Phys. Rev. Lett. 14, 305 (1965).
\bibitem{Bustarret2015} E. Bustarret, Phys. C Supercond. Appl. 514, 36 (2015).
\bibitem{Takada1980} Y. Takada, J. Phys. Soc. Japan 49, 1267 (1980).
\bibitem{Ekimov2004} E. A. Ekimov, V. A. Sidorov, E. D. Bauer, N. N. Mel'nik, N. J. Curro, J. D. Thompson, and S. M Stishov, Nature \textbf{428}, 542 (2004).
\bibitem{Hor2010} Y. S. Hor, A. J. Williams, J. G. Checkelsky, P. Roushan, J. Seo, Q. Xu, H. W. Zandbergen, A. Yazdani, N. P. Ong, and R. J. Cava, Phys. Rev. Lett. 104, 057001 (2010).
\bibitem{Zhang2015} X.-L. Zhang and W.-M. Liu, Sci. Rep. 5, 8964 (2015).
\bibitem{DasSharma2013} S. Das Sarma and Q. Li, Phys. Rev. B 88, 081404 (2013).
\bibitem{Li2014} D. Li, B. Rosenstein, B. Y. Shapiro, and I. Shapiro, Phys. Rev. B 90, 054517 (2014).

\bibitem{Butch2011} N. P. Butch, P. Syers, K. Kirshenbaum, A. P. Hope, and J. Paglione, Phys. Rev. B 84, 220504 (2011).
\bibitem{Bay2012} T. V. Bay, T. Naka, Y. K. Huang, and A. de Visser, Phys. Rev. B 86, 064515 (2012).
\bibitem{Bay2014} T. V. Bay, M. Jackson, C. Paulsen, C. Baines, A. Amato, T. Orvis, M. C. Aronson, Y. K. Huang, and A. de Visser, Solid State Commun. 183, 13 (2014).

\bibitem{QE} P. Giannozzi, S. Baroni, N. Bonini, M. Calandra, R. Car, C. Cavazzoni, D. Ceresoli, G. L. Chiarotti, M. Cococcioni, I. Dabo, A. Dal Corso, S. de Gironcoli, S. Fabris, G. Fratesi, R. Gebauer, U. Gerstmann, C. Gougoussis, A. Kokalj, M. Lazzeri, L. Martin-Samos, N. Marzari, F. Mauri, R. Mazzarello, S. Paolini, A. Pasquarello, L. Paulatto, C. Sbraccia, S. Scandolo, G. Sclauzero, A. P. Seitsonen, A. Smogunov, P. Umari, and R. M. Wentzcovitch, J. Phys. Condens. Matter 21, 395502 (2009).

\bibitem{DalCorso2010} A. Dal Corso, Phys. Rev. B 82, 075116 (2010).
\bibitem{DalCorso2014} A. Dal Corso, Comp. Mat. Sci. 95, 337 (2014).

\bibitem{elk} http://elk.sourceforge.net

\bibitem{Wierzbowska2005} M. Wierzbowska, S. de Gironcoli, P. Giannozi, arXiv:cond-mat/0504077v2

\bibitem{Margine2013} E. R. Margine and F. Giustino, Phys. Rev. B 87, 024505 (2013).
\bibitem{Allen1975} P. B. Allen and R. C. Dynes, Phys. Rev. B 12, 905 (1975)

\bibitem{Takada1993} Y. Takada, J. Phys. Chem. Solids 54, 1779 (1993).

\bibitem{Pagliuso1999} P. G. Pagliuso, C. Rettori, M. E. Torelli, G. B. Martins, Z. Fisk, J. L. Sarrao, M. F. Hundley, and S. B. Oseroff, Phys. Rev. B 60, 4176 (1999).

\bibitem{Oguchi2001} T. Oguchi, Phys. Rev. B \textbf{63}, 125115 (2001).

\bibitem{Mackenzie2003} A. P. Mackenzie and Y. Maeno, Rev. Mod. Phys. 75, 657 (2003).

\bibitem{Graf2011} T. Graf, C. Felser, and S. S. P. Parkin, Prog. Solid State Chem. 39, 1 (2011).
\end{thebibliography}
\end{document}